\documentclass[letterpaper,12pt]{JHEP3}
\usepackage{graphicx}
\usepackage{latexsym}
\newlength{\dummysp}
\newcommand{\beq}{\begin{eqnarray}}
\newcommand{\eeq}{\end{eqnarray}}

\newcommand{\stxt}[1]{\mathop{\hbox{{\scriptsize #1}}}\nolimits}
\newcommand{\bbar}[1]{{\overline{#1}}}
\newcommand{\half}{{1 \over 2}}
\newcommand{\third}{{1 \over 3}}

\newcommand{\beqa}{\begin{eqnarray}}
\newcommand{\eeqa}{\end{eqnarray}}
\newcommand{\nnn}{ \nonumber \\ }

\newcommand{\p}{{\partial}}

\newcommand{\e}{{\epsilon}}
\newcommand{\s}{{\sigma}}
\newcommand{\vev}[1]{{\langle #1 \rangle}}

\newcommand{\ord}[1]{{{\cal O}(#1)}}
\newcommand{\gappeq}{\mathrel{\rlap {\raise.5ex\hbox{$>$}}
{\lower.5ex\hbox{$\sim$}}}}
\newcommand{\lappeq}{\mathrel{\rlap{\raise.5ex\hbox{$<$}}
{\lower.5ex\hbox{$\sim$}}}}
\newcommand{\myref}[1]{(\ref{#1})}

\newcommand{\ben}{\begin{enumerate}}
\newcommand{\een}{\end{enumerate}}
\newcommand{\bit}{\begin{itemize}}
\newcommand{\eit}{\end{itemize}}

\newcommand{\Sb}{\bbar{S}}

\newcommand{\sbb}{\bbar{s}}
\newcommand{\Lcal}{{\cal L}}
\newcommand{\thb}{\bbar{\theta}}
\newcommand{\Dc}{\bbar{D}}
\newcommand{\hc}{{\rm h.c.}}
\newcommand{\fourth}{{1 \over 4}}
\newcommand{\Ub}{\bbar{U}}
\newcommand{\Wb}{\bbar{W}}
\newcommand{\ddd}{\nnn &&}
\newcommand{\Sigb}{\bbar{\Sigma}}
\def\[{\left [}
\def\]{\right ]}
\def\({\left (}
\def\){\right )}
\def\|{|_{(0,0)}}
\title{Instanton effects and linear-chiral duality}

\author{Joel Giedt  \\ Department of Physics, University of Toronto \\
60 Saint George Street, Toronto, ON M5S 1A7 Canada \\
\email{giedt@physics.utoronto.ca}}

\author{Brent D. Nelson \\
Michigan Center for Theoretical Physics, University of Michigan \\
3444A Randall Laboratory, Ann Arbor, MI 48109 USA \\
\email{bdnelson@umich.edu}}

\preprint{hep-th/0307224 \\ MCTP 03-35 \\ May 18, 2004}

\abstract{We discuss duality between the linear and chiral dilaton
formulations, in the presence of super-Yang-Mills
instanton corrections to the effective action.  
In contrast to previous work on the subject, our approach appeals
directly to explicit instanton calculations and does not
rely on the introduction of an auxiliary Veneziano-Yankielowicz
superfield.  We discuss duality in the
case of an axion that has a periodic scalar potential,
and find that the bosonic fields of the dual 
linear multiplet have a modified interpretation.
We note that symmetries of the axion potential
manifest themselves as symmetries of the equations
of motion for the linear multiplet.  We also make
some brief remarks regarding dilaton stabilization.
We point out that corrections
recently studied by Dijkgraaf and Vafa can be used to stabilize the
axion in the case of a single super-Yang-Mills condensate.}


\begin{document}

\section{Introduction}
Quite impressive and reliable results have been obtained for the
instanton generated nonperturbative superpotential in
super-Yang-Mills (SYM) and super-QCD
\cite{ADS}.  These results have been further refined
by computations of corrections due to decoupled matter,
sparked by recent work \cite{DV02,Cac02} that is currently
the subject of intense interest and activity.

Here we discuss the duality between the dilaton described
by a linear multiplet \cite{linOR} $L$ and
the dilaton described by a chiral multiplet $S$,
in rigid N=1 4d supersymmetry.
While this has been discussed at length with regard
to anomaly cancellation \cite{anoc},
in the present article we aim to describe
instanton effects in the dual formalism.
Such effects play a crucial role in string-inspired
models of moduli stabilization; for example, \cite{BGW}.
In contrast to previous work, such
as in \cite{anoc,BGW}, we discuss this duality 
without introducing a
Veneziano-Yankielowicz (VY) superfield \cite{VY82}.
This is an auxiliary superfield that produces
the known instanton superpotential when it is
integrated out.\footnote{The VY superfield can also be regarded as
a background field of the 2PI effective action \cite{Bur95}.}
We avoid the VY superfield because we would prefer to
understand the duality without ever ``integrating in''
this superfluous field in the first place.\footnote{
In Section \ref{dvsec}, we will make some remarks
in regard to the superpotential corrections computed by
Dijkgraaf and Vafa.  At that juncture it will be
convenient to make use of the VY superfield in order
to make contact with their notation.}

We now summarize the content of our paper and
our key results:
\bit
\item
In Section \ref{etm} we briefly describe the motivations
for studying nonperturbative corrections to the
dilaton potential:  first because they are certainly
present, as a consequence of instanton configurations, and second
because these corrections can play an important role
in stabilizing the vacuum.
\item
Using straightforward manipulations in the superfield formalism,
we find in Section \ref{scld} that when the
linearity of the multiplet $L$ is sufficiently modified,
this formalism is exactly equivalent to the one
involving the chiral multiplet $S$.
\item
We have verified our results by also performing the duality transformation
at the component field level.  In Section \ref{cfm} we
discuss the translation between the two formalisms in terms
of component fields.
\item
In Section \ref{abr}, we address an apparent
inconsistency between the two formalisms which
has appeared in the literature \cite{BGW}.  We point
out a simple calculus error that was made,
which when corrected, resolves the apparent difficulty
for exact duality.
\item
In Section \ref{adua} we study the equations of motion
for the bosonic fields in the modified linear multiplet.
We find that the traditional interpretation of the 1-form that is
contained in the linear multiplet---as the Hodge dual of a
field strength for a 2-form---is modified if the axion
has a potential.
\item
We find that symmetries of the axion potential are
re-expressed in the dual formalism as symmetries of the
equations of motion for the 1-form.  This is also discussed
in Section \ref{adua}.
\item
We point out that the corrections computed by Dijkgraaf and Vafa
\cite{DV02} may be used to stabilize the axion using a
single SYM condensate.  Details on this matter may
be found in Section \ref{dvsec}.
\item
In Section \ref{outl} we give our concluding remarks
and comment on issues for further research.
\eit

\section{Effective theory and motivations}
\label{etm}
In this section we review certain well-known facts regarding instanton
corrections in pure SYM.  Here, the pure SYM theory must be understood
as resulting from a more fundamental theory with $N_f=N_c-1$ flavors
of fundamental matter, where $N_c$ is the number of 
colors; i.e., $SU(N_c)$ super-QCD.  As is well-known, instantons
generate a nonperturbative superpotential in the $N_f=N_c-1$
theory \cite{ADS}.  One can then obtain an effective pure SYM,
together with a nonperturbative superpotential, valid below a
scale $\mu$, by studying the theory along a flat direction where 
all of the flavors obtain masses of order $\mu$.  For a review
and citations to the original literature, we refer the reader
to \cite{Intriligator:1995au,Peskin:1997qi}.

We will work in the chiral dilaton formulation, 
where these results are most familiar.  
The scalar component $s$ of the chiral dilaton field $S$
determines the effective gauge
coupling and the effective theta angle at a scale $\mu$ through
its vacuum expectation value ({\it vev}):
\beq
\vev{s} = \frac{1}{g^2} - i \frac{\theta}{8 \pi^2} ~.
\eeq
In the case of pure SYM the effective superpotential for $S$
is expressed in terms of a superfield extension of the ordinary
dynamical YM scale:
\beq
\Lambda = \mu \exp \( - 8 \pi^2 S/b \), \qquad
W(S) = \tilde c \Lambda^3 ,
\label{suop}
\eeq
where for example $b=3N_c$ for pure $SU(N_c)$.
Generally we write the K\"ahler potential as
\beq
K(S + \Sb) = \mu^2 k(S+\Sb).
\eeq
In examples we will examine specific forms for $k(S+\Sb)$.
We always assume that $K$ is a
function of $S + \Sb$, but not of $S - \Sb$.
Consequently, $\p K/\p S = \p K / \p \Sb
= \p K / \p(S + \Sb)$,
or in a more abbreviated notation $K_S = K_\Sb = K'$.

Quite often in the literature on supergravity and string-inspired
effective theories, only the leading order K\"ahler potential,
$-\mu^2 \ln(S+\Sb)$, is used; however, just
as the superpotential receives instanton corrections,
the K\"ahler potential will likewise be modified
by nonperturbative effects.  Due to a lack of holomorphy, it is difficult
to obtain any reliable information on the form of
$K$ in the nonperturbative regime.  However, instanton
corrections are certainly present.
For example, if we start from the leading order $k = -\ln(S+\Sb)$, 
and the nonperturbative superpotential \myref{suop},
the 1-loop corrections to the K\"ahler potential
take the form
\beq
\delta K \propto (k'')^{-2} |W''|^2 
\propto (S + \Sb)^4 (\Lambda \bbar{\Lambda})^3
\label{kerl}
\eeq
Higher orders in perturbation theory and nonperturbative
effects will generalize this result in a way that
we cannot presently determine.  For example, nonperturbative
corrections might allow for integral powers of $\Lambda \bbar{\Lambda}$
that are not multiples of 3, since they would not necessarily
derive from the superpotential, as was the case with \myref{kerl}.
We can, however, be
confident that the leading order K\"ahler potential
is just an approximation that in some regimes may
prove to be inaccurate.

Rather than ambling along
with a form of $K$ that ignores this reality, we find
it more logical to explore various ``reasonable'' forms
for the nonperturbative K\"ahler potential; and, to classify
the qualitative results that follow.  Indeed, intuition leads
us to believe that some {\it ad hoc} assumptions are better motivated
than others.

As an example, we find it entirely sensible to
include instanton effects in the
K\"ahler potential in the most naive way,
based on dimensional analysis:
\beq
K(S + \Sb) = \mu^2 k(S+\Sb)
= -\mu^2 \ln(S+\Sb) + c(S+\Sb) \Lambda \bbar{\Lambda}
+ \ord{\Lambda^2 \bbar{\Lambda}^2/\mu^2} ,
\label{jhah}
\eeq
with $c(S+\Sb)$ a slowly varying function of $S+\Sb$ and the
$\ord{\Lambda^2 \bbar{\Lambda}^2/\mu^2}$ terms presumably
negligible.  Of course more general assumptions exist; say,
fractional powers of $\Lambda \bbar{\Lambda}$ appearing in \myref{jhah},
or functions of $\Lambda + \bbar{\Lambda}$.
Such generalizations
are captured by allowing $k(S+\Sb)$ to be arbitrary.

Our interest in nonperturbative corrections to $K$ is not
just academic.
In rigid supersymmetry, the scalar potential
that determines the vacuum is given by
\beq
V_{\stxt{rigid}} &=& \mu^4 \frac{|24 \pi^2 \tilde c|^2}{b^2 k''(s + \sbb)}
\exp\[-\frac{24 \pi^2}{b} (s + \sbb)\] \nnn
&=& \mu^4 \frac{|\tilde c|^2 }{\p_x^2 k(x)} \exp(-x),
\qquad x =\frac{24 \pi^2}{b} (s + \sbb) .
\eeq
Generally, the effect of corrections such as assumed in
\myref{jhah} is merely\footnote{We do not consider
corrections that are radically different from
\myref{jhah}; for example, in \cite{Arkani-Hamed:1998nu} it was found
that a form of $K(S+\Sb)$ can be engineered to
yield a nontrivial minimum in the rigid susy case.
Our results for chiral-linear duality, however,
may also be applied to this $K(S+\Sb)$.}
to slightly shift the location
and height of the maximum of $V_{\stxt{rigid}}$.
As an example, we consider the case where the function
$c(s + \sbb)$ amounts to what is essentially a
polynomial in $g^2$:
\beq
k(x) = \ln(24 \pi^2/b) - \ln(x) + (c_1 + c_2 x^{-1}) \exp(-x/3) .
\label{kass}
\eeq
(The constant term is
due to the replacement of $s + \sbb$ with
$x$; it is irrelevant here but must be kept track of
for supergravity considerations below.)
Eq.~\myref{kass} is just the first two terms in \myref{jhah} with
$c(s+\sbb) \propto c_1 +  c_2 b /24 \pi^2 (s+\sbb)$,
with constants $c_1$ and $c_2$.
Thus, the correction behaves like
\beq
\delta K \sim \(c_1 + c_2 \frac{b g^2}{48 \pi^2} \)
\exp \(-\frac{8\pi^2}{bg^2} \) .
\eeq
We believe this to be a reasonable assumption, though
we have no inkling about the magnitude of $c_1$ or
$c_2$ (except that the naive loop factor
has been scaled out of $c_2$, so that this
sort of suppression does not seem implied for $c_2$).  
Higher orders in $g^2$ certainly will appear
in $c(s + \sbb)$, but for our illustration, which is
only meant to be qualitative, we neglect them
as small.  For convenience, we
define $\hat V_{\stxt{rigid}}(x)
= V_{\stxt{rigid}}(x)/\mu^4 |\tilde c|^2$.
In Fig.~\ref{f1} the dashed line shows
this quantity as a function of $x$ for the
case of $c_1=0$ and $c_2=1$.
It can be
seen that the well-known runaway behavior to vanishing and
infinite couplings is retained.

On the other hand if we generalize to supergravity, the scalar potential
reads instead
\beq
V_{\stxt{sugra}} &=& \mu^4 |\tilde c|^2 e^{-x+k(x)}
\[ (\p_x^2 k(x))^{-1} (1-\p_x k(x))^2 -3 \] .
\label{supo}
\eeq
The supergravity corrections have important effects
given the assumed instanton-induced corrections
\myref{kass}.  For convenience, we define $\hat V_{\stxt{sugra}}(x)
= 24 \pi^2 V_{\stxt{sugra}}(x)/b \mu^4 |\tilde c|^2$.
In Fig.~\ref{f1} the dotted line shows
this quantity as a function of $x$,
again for the case of $c_1=0$ and $c_2=1$.
In the regime where $x^{-1} \exp(-x/3)$ competes
with the ``leading order,'' i.e. where $x \leq \ord{1}$,
the modification is so great as to create a deep minimum---and,
to completely lift the infinite coupling runaway.

Clearly, the $g^4 \sim x^{-2}$ corrections to $c(s + \sbb)$
that have been neglected in \myref{kass} will
modify the details of the strong coupling behavior.
It is also true that in this same regime 
$\Lambda \bbar{\Lambda}/\mu^2 = \exp(-x/3) = \ord{1}$,
so the truncation of \myref{jhah} is not justified.
Nevertheless, as higher powers of $\Lambda \bbar{\Lambda}/\mu^2$
are added in, it is a generic feature that the strong
coupling runaway behavior tends to be removed and nontrivial
(typically local) minima are created.

Note that for $c_1=0$ and $c_2=1$ the vacuum energy of is order $-\mu^4$;
it is well-known that further refinements and
fine-tuning can be used to manipulate this
so-called {\it K\"ahler stabilization} of the dilaton \cite{kahstab,BGW}.
Indeed, it is possible to obtain an
approximately vanishing cosmological constant and 
to stabilize at a ``weak coupling'' minimum.
As an example, we have fixed $c_2=1$ and
then tuned $c_1$ to a value such that the minimum occurs at vanishing
cosmological constant.  We find that $c_1=3.27$
works well; the corresponding potential is indicated in Fig.~\ref{f1}
by the solid line.  Naturally the nonperturbative
corrections to the K\"ahler potential do not alter
the asymptotic behavior at weak coupling, where they
are totally negligible; the weak coupling runaway
persists.  However, with the cosmological constant
tuned to zero the barrier height is considerable:
$E_{\stxt{barrier}} \approx 10 \mu (b |\tilde c|^2/24 \pi^2)^{1/4}$.
If $\mu$ is of order the 4d Planck scale, $2.4 \times 10^{18}$ GeV,
and if $|\tilde c|$ is not too small, the effects of the
approximately degenerate vacuum at $x \to \infty$ would
presumably be neglible, for suitable cosmological
initial conditions.

\FIGURE{
\includegraphics[height=3.0in,width=5.0in]{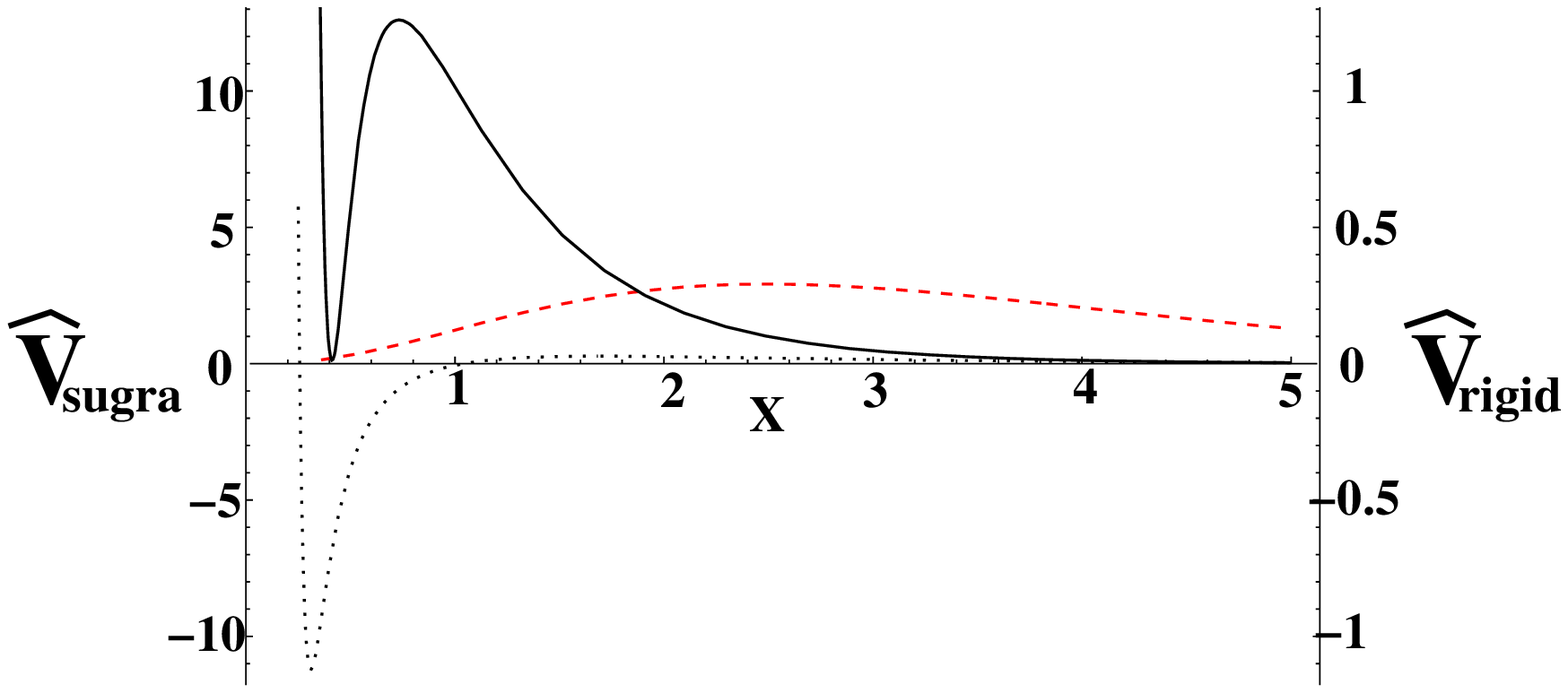}
\caption{A comparison of the effects of nonperturbative
corrections to $K$ for the rigid supersymmetry
and supergravity scalar
potentials, as a function of $x \propto s+\sbb$.
The dashed line represents the rigid supersymmetry potential
for $c_1=0,c_2=1$, while the dotted line is the 
supergravity potential for the same choice of parameters.
On the other hand, the solid line represents the supergravity
potential with $c_1=3.27$ and $c_2=1$; this yields a
vanishing cosmological constant.}
\label{f1}
}

While the supergravity potential \myref{supo}, together
with assumptions for $K$ such as \myref{kass}, may be
used to stabilize $s + \sbb$, it can be seen that the
axion $(s - \sbb)/i$ is absent and thus remains a flat
direction.  To stabilize the axion requires slightly more
complicated assumptions about the form of the effective theory.  
As a very well-known example,
suppose the relevant gauge group has 2 simple factors
with $b_1 \not= b_2$.  Eq.~\myref{suop} is generalized to
\beq
\Lambda_1 = \mu \exp \( - 8 \pi^2 S/b_1 \), \quad
\Lambda_2 = \mu \exp \( - 8 \pi^2 S/b_2 \), \quad
W(S) = \tilde c_1 \Lambda_1^3 + \tilde c_2 \Lambda_2^3. \quad
\label{suoq}
\eeq
$V$ is no longer just a function of $s + \sbb$.
In this case the axion $(s - \sbb)/i$ is also stabilized.
Another possibility is to add heavy matter that has
a nontrivial vacuum.  When this matter is integrated
out, as will be seen in Section \ref{dvsec}, it is possible
to obtain corrections to \myref{suop} that stabilize
the axion.

In summary, instanton effects---however they might
appear in the K\"ahler potential---play an important role
in the stabilization of the dilaton and axion.
Thus it may be of interest to translate between
the chiral dilaton formulation and the linear dilaton
formulation when these effects are present.  Furthermore,
we would like to be able to do so without introducing
additional machinery, such as the Veneziano-Yankielowicz
auxiliary superfield.  In the next section we elucidate
how this is to be done in the case of rigid supersymmetry.

\section{Linear-chiral duality}
\label{scld}
We now want to address these instanton corrections in the
the dual linear dilaton formulation.
To do this we want to begin with the chiral dilaton
formulation and translate to an equivalent system;
this is the so-called {\it duality transformation.}
It is essential that field redefinitions are made
that respect both:  (i) the equations of motion,
and (ii) any constraint equations.  To this end we write
a first-order Lagrangian whose equations of motion
contain both (i) and (ii):  the constraints are
imposed dynamically.

We begin with the dilaton effective Lagrangian in the chiral formulation,
written in superspace notation:\footnote{Our conventions
are those of \cite{WB}.}
\beq
\Lcal = \int d^4\theta \; K(S+\Sb) + \[ \int d^2 \theta \;
W(S) + \hc \] ~.
\label{slag}
\eeq
We replace this with a first-order Lagrangian that,
as will be shown below, imposes
the chirality constraints, using Lagrange multiplier
superfields $U,\Ub$:
\beq
\Lcal_{FO} &=& \int d^4\theta \; K(S+\Sb) \ddd
+ \int d^4\theta \[ U (S - \fourth \Dc^2 \Sigma)
+  \Ub (\Sb - \fourth D^2 \Sigb) \] \ddd
+ \int d^2 \theta \; W(\fourth \Dc^2 \Sigma)
+ \int d^2 \thb \; \Wb(\fourth D^2 \Sigb) ~.
\label{wuyr}
\eeq
Note that in \myref{wuyr} all superfields 
$S,\Sb,U,\Ub,\Sigma,\bbar{\Sigma}$ are unconstrained.

The simplest superfield equations of motion occur for fields that appear
only in D-density terms; i.e., only under $\int d^4 \theta$.
The superfield equations of motion obtained from
varying $U$ and $\Ub$ yield the chirality and antichirality constraints:
\beq
S = \fourth \Dc^2 \Sigma ~, \qquad \Sb = \fourth D^2 \bbar{\Sigma} ~.
\label{ssig}
\eeq
Varying with respect to $S$ and $\Sb$ yields simply
\beq
0=U+K'(S+\Sb)=\Ub+K'(S+\Sb) ~.
\eeq
Thus when we impose the equations of motion
that follow from \myref{wuyr},
we obtain the on-shell projection to a real multiplet $L$:
\beq
L = U = \Ub = -K'(S+\Sb) ~.
\label{ldff}
\eeq

Note that if we had just used $L$ in place
of $U$ and $\Ub$ from the start, we
would not have the two equations in~\myref{ssig} independently.
Instead, the equations of motion would only require
\beq
S + \Sb = \fourth \( \Dc^2 \Sigma + D^2 \bbar{\Sigma} \) .
\label{tsig}
\eeq
Since $S$ and $\Sb$ are unconstrained superfields in the first
order Lagrangian \myref{wuyr}, the constraint \myref{tsig} does
not enforce (anti-)chirality by its equations of motion.
Of course,
the identification \myref{ssig} is {\it a particular} solution
to \myref{tsig}.  But for the duality
to be faithful, the equations of motion must have
\myref{ssig} as a {\it unique} solution.
In this respect our duality transformation is
more restrictive than the one that has previously
been imposed in the literature \cite{anoc,BGW}.

While \myref{ldff} implicitly tells us how to replace $S+\Sb$
with $L$ in the Lagrangian, the superpotential terms
will involve $S$ and $\Sb$ separately.  
This is particularly important where the
axion has a potential.  In the following
manipulations we will see that the necessary data is
obtained from variation of \myref{wuyr} with respect to $\Sigma$ and $\Sigb$,
which leads to modified linearity conditions for $L$.
Varying with respect to $\Sigma$ and $\Sigb$
requires that we handle a mixture of D-density and
F-density terms.
To perform the analysis we rewrite the D-density terms
that contain these fields using
\beq
\int d^4 \theta U \Dc^2 \Sigma =
- \fourth \int d^2 \theta \Dc^2 \( U \Dc^2 \Sigma \)
\label{koko}
\eeq
and similarly for the $D^2 \Sigb$ term.\footnote{See for
example p.~67 of \cite{WB}.}

We vary $\Sigma$ to obtain
\beq
\delta \Lcal_{FO}
&=& {1 \over 16} \int d^2 \theta \[
\Dc^2 \( U \Dc^2 \delta \Sigma \) + 4 W'(\fourth \Dc^2 \Sigma)
\Dc^2 \delta \Sigma \] \nnn
&=& {1 \over 16} \int d^2 \theta \[
\Dc^2 \( \Dc^2 U \delta \Sigma \) + 4 W'(S)
\Dc^2 \delta \Sigma \] \nnn
&=& -{1 \over 4} \int d^4 \theta \[ \delta \Sigma
\( \Dc^2 U + 4 W'(S) \) \] .
\eeq
In the second line we use the equations of
motion \myref{ssig} and the identity
\beq
0 = \Dc^2 \[ \Dc^2 U \delta \Sigma - U \Dc^2 \delta \Sigma \] .
\eeq
In the 3rd line we use $\Dc^2 S \propto \Dc^2 \Dc^2 \Sigma = 0$
and reverse the type of manipulation that led to \myref{koko}.
A similar analysis leads us to write the variation
with respect to $\Sigb$ as:
\beq
\delta \Lcal_{FO}
&=& -{1 \over 4} \int d^4 \theta \[ \delta \Sigb
\( D^2 \Ub + 4 \Wb'(\Sb) \) \] .
\eeq

Vanishing of these two variations leads to the constraints:
\beq
\Dc^2 U + 4 W'(S) = 0 ,
\qquad
D^2 \Ub + 4 \Wb'(\Sb) = 0 .
\eeq
Taking into account \myref{ldff} we arrive at the modified
linearity conditions
\beq
\Dc^2 L = - 4 W'(S), \qquad
D^2 L = - 4 \Wb'(\Sb).
\label{moli}
\eeq
These, together with \myref{ldff},
are sufficient to (implicitly) redefine the components of $S,\Sb$
in terms of the components of $L$.

\section{Component fields}
\label{cfm}
We have verified all of the above superfield relations
at the level of component fields.  This straightforward
exercise begins by writing out the unconstrained superfields
appearing in \myref{wuyr} in terms of $\theta,\thb$ expansions.
For example, keeping only bosons, the real part of $U$ is
given by
\beq
L = \ell + \theta \s^m \thb h_m
+ \theta^2 Z + \thb^2 \bbar{Z} + \theta^2 \thb^2 (D + \fourth \Box \ell).
\label{lexp}
\eeq
Recall that at this point we want $L$ to be unconstrained,
except for the fact that it is defined as $L = (U + \Ub)/2$.
Thus $L$ contains independent fields $Z,\bbar{Z}$ and the
1-form $h^m$ is completely general.  These degrees of
freedom will be subject to constraints---i.e., the modified
linearity conditions---when we go on shell.
It is straightforward to substitute 
analogous expressions for the unconstrained
superfields into \myref{wuyr} and
to work out the component expansion.  From there,
one may work out the equations of motion.
There are many, and we will spare
the details.  Here we will just state key results.

The Lagrangian dual to \myref{slag}, neglecting fermions,
is merely:
\beq
&& \Lcal(\ell,h^m) = -\fourth \ell ~ \Box ~
\[ s(\ell,\p_m h^m)+\sbb(\ell,\p_m h^m) \] \nnn
&& \qquad - {1 \over K''(s(\ell,\p_m h^m)+\sbb(\ell,\p_m h^m))}
\( \fourth h^m h_m + |W'(s(\ell,\p_m h^m))|^2 \) .
\label{lagL}
\eeq
Here we have shown that $s=s(\ell,\p_m h^m)$ wherever it appears.
It remains to specify how this is obtained.

Implicitly, we can replace $s+\sbb$ everywhere using
\beq
\ell = -K'(s+\sbb) .
\eeq
Implicitly, we can replace $s-\sbb$ using the additional constraint
\beq
\p^m h_m = {i \over K''(s+\sbb)} \(W'(s) \Wb''(\sbb)
- \Wb'(\sbb) W''(s) \) ~.
\label{phmm}
\eeq
Up to some factors that involve $s+\sbb$,
one sees that $\p^m h_m$ is identified with $V'(a)$,
the axion force term obtained from the potential $V(a)$ for the
axion $a=(s-\sbb)/i$.
This is not surprising, because we have the $\theta \thb$
part of the constraint \myref{ldff}:
\beq
h^m = {1 \over i} K''(s+\sbb) \p^m(s-\sbb)
\eeq
Thus $\p^m h_m$ should involve $\Box a$.  But the
equations of motion for the axion relate $\Box a$ to $V'(a)$.

For the $\theta^2$ and $\thb^2$ components of $L$ we have
\beq
Z=L|_{\theta^2} = \Wb'(\sbb), \qquad \bbar{Z}=L|_{\thb^2} = W'(s).
\label{auxe}
\eeq
This is the $\theta=\thb=0$ part of the modified linearity conditions
\myref{moli}.
In fact it is straightforward to check that \myref{moli}
is consistent with the component field equations of motion to all orders
in $\theta, \thb$.  We remark that in the presence of the
instanton effects, the modified linear multiplet contains
auxiliary fields.  In particular, \myref{auxe} indicates
that $L$ contains the F-term component of the chiral
dilaton and its conjugate, $F_S$ and $F_\Sb$.
This, of course, has been noted before in formalisms
that relied on a Veneziano-Yankielowicz 
superfield \cite{anoc,BGW,Binetruy:2000md}.

\section{A brief remark}
\label{abr}
We note in passing that an apparent inequivalence
between the two formalisms was noted in the Appendix of \cite{BGW}.
There it was found in the linear dilaton formalism that
the {\it K\"ahler moduli} $t^I$ of a string-inspired effective
supergravity were stabilized at the {\it self-dual} values
(with respect to an $SL(2,Z)$ isometry of the scalar
manifold) of $1$ or $e^{i\pi/6}$.
A duality transformation was made in the
Appendix of \cite{BGW}, and it was found that ``the minimum is shifted
slightly away'' [from the self-dual value].
However, we find that the apparent conflict with linear-chiral
duality is resolved by noting a simple error that was
made by the authors of \cite{BGW} in obtaining their
Eq.~(A.16) from their Eq.~(A.14).
They have kept too many quantities
constant in performing the differentiation---simultaneously both
$s + \sbb$ and $\ell$.
But since $t^I + \bbar{t}^I$ mixes with $s + \sbb$ to
give $\ell$, in their Eq.~(A.1), this is not right.
Once the chain rule is properly applied, it is not hard
to show that their dual chiral formulation also predicts
stabilization of $t^I$ at the self-dual values.
(We do not provide further details because they just
involve elementary manipulations.  However, we thought
it important to resolve this apparent problem for
lineal-chiral duality.)

\section{Dual description for the axion}
\label{adua}
If we have only one condensate and a superpotential
of the form \myref{suop}
then $|W'(s)|^2$ is independent
of $s-\sbb$; the axion has no potential and is massless.
Equivalently, the right-hand side of \myref{phmm}
vanishes identically.
This constraint equation has the general solution
\beq
h^m = \e^{mnpq} \p_n b_{pq} .
\label{tffs}
\eeq
Thus, as has been known for a very long time,
the 1-form in the linear multiplet is Hodge dual
to a 2-form field strength \cite{linOR}.

In more general situations the axion gets a potential
from the instanton physics.  In this case the right-hand side of
\myref{phmm} does not vanish.  Eq. \myref{tffs} is inconsistent
with the constraints.  The 1-form in the linear multiplet
must be reinterpreted.  It is no longer just the Hodge dual of
a 2-form field strength.  Instead, it is the Hodge dual of
a massive 3-form \cite{AT81}.

\subsection{Massive axion dual}
Temporarily we oversimplify and consider just
an ``axion'' with a constant, field independent mass.
Thus we assume
\beq
\Lcal(a) = - \half \p_m a \p^m a - \half m^2 a^2 .
\eeq
The duality is obtained with the identification
\beq
h^m \equiv \p^m a \quad \Rightarrow \quad \p^m h_m = \Box a = m^2 a ,
\label{deaa}
\eeq
where in the second step we use the equation of motion for $a$.
This equation of motion and the constraint (relating the 1-forms
$h_m$ and $\p_m a$) are obtained from the first order Lagrangian
\beq
\Lcal_{FO} = \half \p_m a \p^m a - \half m^2 a^2 + h^m \(
h_m - 2 \p_m a \) .
\eeq
It is easy to check that the dual theory obtained by eliminating
$a$ from $\Lcal_{FO}$ through {\it its} equations of motion
is given by
\beq
\Lcal(h_m) =  - {1 \over 2 m^2} \( m^2 h^m h_m + \p^m h_m \p^n h_n\) .
\eeq
The equations of motion that follow from $\Lcal(h_m)$ are
\beq
\p^m \p^n h_n - m^2 h^m = 0 .
\label{heom}
\eeq
The general solution is nothing but \myref{deaa}.  Thus, {\it the
axion just parameterizes the general solution to the 1-form
equations of motion.}  We will see that this is likewise true
in the more interesting circumstance of the 1-form dual to
an {\it interacting} axion.

Note that \myref{heom} is not the usual
equation of motion for a massive vector boson.
In Fourier space the mode expansion coefficients $a_{{\bf p}}^m$
of $h^m$ are not independent.
Instead they satisfy
\beq
a_{{\bf p}}^m = {p^m \over \sqrt{{\bf p}^2 + m^2}} a^0_{{\bf p}} ~.
\label{modr}
\eeq
In the rest frame, the spatial components of $h^m$ vanish;\footnote{This
can also be seen directly from \myref{heom}.} there is only
1 on shell degree of freedom.
From this we understand how the 1-form can be equivalent to a 0-form.
In fact, this is precisely the behavior of a massive 3-form \cite{AT81}.

\subsection{Axion with periodic potential}
Here we suppose
\beq
\Lcal(a) = -\half \p_m a \, \p^m a + m^4 \cos \( {a \over m} \) .
\label{ahrt}
\eeq
The degenerate vacua are labeled by an integer $n$,
indicating $\langle n | a | n \rangle = 2 \pi n$.  The mass of a fluctuation
about any of these vacua is $m$.  It is interesting to see
how this circumstance is reflected in the dual 1-form theory.
To this end, we write down an equivalent, first order Lagrangian:
\beq
\Lcal_{FO} = \half \p_m a \p^m a + m^4 \cos \( {a \over m} \)
+ h^m \( h_m - 2 \p_m a \) .
\eeq
The equations of motion obtained from $\Lcal_{FO}$ are
\beq
h^m = \p^m a, \qquad
0 = \Box a + m^3 \sin \( {a \over m} \) - 2 \p_m h^m,
\label{ahru}
\eeq
clearly equivalent to the equation of motion that follows
from \myref{ahrt}.
Next we eliminate $a$ to obtain the equivalent
1-form Lagrangian.  Differentiating the second
equation in \myref{ahru} and contracting with $h^n$
it is easy to show
\beq
m^4 \cos \( {a \over m} \) =  m^2 {h^n \p_n \p_m h^m \over
h_p h^p } .
\eeq
With this result, one finds
\beq
\Lcal(h) = - \half h^m h_m + m^2 {h^n \p_n \p_m h^m \over
h_p h^p } .
\label{jatw}
\eeq
The equations of motion that follow from \myref{jatw}
are not illuminating, and we need not write them here.
The sole thing worth noting about them is that because
of the duality transformation that has been made, we are
guaranteed that they have a general solution $h^m = \p^m \psi$,
where $\psi$ is an integral of the differential equation
\beq
\Box \psi = m^3 \sin \( {\psi \over m} \) .
\eeq
But this is nothing other than the axion equation of motion.
Thus, {\it the periodicity
of the axion potential is reflected in a degeneracy of solutions
to the equivalent 1-form equations of motion.}  This
degeneracy is not immediately apparent (to us) upon
inspection of \myref{jatw}.  For this reason, the chiral
formulation seems advantageous for understanding
the pseudoscalar vacuum of the theory.


\section{Single condensate stabilization}
\label{dvsec}
We now discuss stabilization of the dilaton
and axion using only a single SYM condensate.
To achieve this, we appeal to the corrections---to
the Veneziano-Yankielowicz (VY) superpotential in the
case where a very heavy adjoint chiral superfield is
present---worked
out recently by Dijkgraaf and Vafa (DV) \cite{DV02}.
We use the DV result that the VY superpotential, which
contains the VY auxiliary superfield $U$, can be written
\beq
W(S,U) = NU \[ \ln {U \over \mu^3} + {8\pi^2 \over N} S + f(U) \] .
\eeq
Here $f$ is a calculable power series in the VY superfield:
\beq
f(U) = \sum_{n>0} c_n \( { U \over \mu^3} \)^n  .
\label{pwsr}
\eeq
The equations of motion for the VY superfield are:
\beq
{\p W(S,U) \over \p U} = 0
= N \[\ln {U \over \mu^3} + {8\pi^2 \over N} S + f(U)
+ U {\p f(U) \over \p U} + 1 \] .
\label{kiur}
\eeq
The solution to \myref{kiur} can be written implicitly as
\beq
U=\Lambda^3 \exp \( - f - U {\p f \over \p U} \), \qquad
\Lambda^3&=&\mu^3 e^{-1} \exp \( -{8\pi^2 \over N} S \) .
\eeq
Note that (up to an unimportant factor $e^{-1}$)
we have retained the leading order definition \myref{suop}
of $\Lambda$, the uncorrected dynamical (``QCD'') scale.
The VY superfield is auxiliary, in that it has no kinetic
term.  It thus has a singular metric, and consequently
an effectively infinite mass.  We will therefore 
eliminate $U$ with its equations of motion.

As an aside, we note that, upon imposing the equation of
motion \myref{kiur},
\beq
\frac{\p^2 W(S,U)}{\p U^2}
= \frac{N}{\Lambda^3} \( 1 + \ord{\Lambda^3/\mu^3} \) .
\eeq
Thus even if $U$ {\it had} a (canonical) kinetic term,
it would have a very large mass $m_U$:
\beq
m_U = \ord{N \mu^3 / \Lambda^3} \cdot \mu \gg \mu .
\eeq
At energies of order $\Lambda^3$ it is essentially
static, and certainly should be integrated out in
discussing the low energy effective theory.\footnote{We
thank Erich Poppitz for emphasizing this to us.}  
The case where the gaugino bilinear has
been treated as a dynamical field has been studied in
much greater detail in \cite{Wu:1996en}, 
where a similar conclusion was reached
about the effective mass scale of this composite
degree of freedom.

Eliminating $U$ with its equations of motion, we obtain:
\beq
W(S) &\equiv& W(S,U(S)) = \left. -N U \(1 + U {\p f(U) \over \p U} \)
\right|_{U(S)} .
\eeq
Provided $\Lambda \ll \mu$ and the coefficients $c_n$
are not unreasonably large (a more precise statement
will be given shortly), we need keep only the leading
term in \myref{pwsr}:
\beq
f \approx c_1 { U \over \mu^3} \approx c_1 {\Lambda^3\over \mu^3} .
\label{thap}
\eeq
The coefficient $c_1$ depends on the massive matter representations
that have implicitly be integrated out to obtain \myref{pwsr}.
As an example we use the DV {\it perturbative}
superpotential with a single adjoint chiral superfield:
\beq
W = \half m \Phi^2 + \third \lambda \Phi^3 \quad
\Rightarrow \quad
c_1 = 2 \lambda^2 \( {\mu \over m} \)^3 .
\label{uurq}
\eeq
We implement the approximation \myref{thap} to obtain:
\beq
U {\p f \over \p U} &\approx& f , \qquad
U \approx \Lambda^3 \exp (-2f) \approx \Lambda^3
\( 1 - 2 c_1 {\Lambda^3\over \mu^3} \),  \nnn
W(S) &\approx& -N\Lambda^3 \( 1 - c_1{\Lambda^3\over \mu^3} \)
= -N\Lambda^3 \( 1 - 2 \lambda^2 \frac{\Lambda^3}{m^3} \).
\eeq
Now we can be more specific about the smallness of corrections:
we require $m^3 \gg \lambda^2 \Lambda^3$ in order that the
approximations be valid.  {\it It is worth emphasizing that it
is not enough to have $\Lambda \ll \mu$.}

To understand the vaccum in the presence of the DV corrections,
note that
\beq
{\p \Lambda^3 \over \p S} = -{8\pi^2 \over N} \Lambda^3 .
\eeq
Thus
\beq
W'(S) \approx 8\pi^2 \Lambda^3 \(1 - 2c_1{\Lambda^3\over \mu^3} \)
= 8\pi^2 \Lambda^3 \( 1 - 4 \lambda^2 \frac{\Lambda^3}{m^3} \) .
\label{jhta}
\eeq
This theory has a supersymmetric vacuum, and therefore
a global minimum, at
\beq
2c_1 \Lambda^3 \approx \mu^3 \quad
\Leftrightarrow \quad
m^3 \approx 4 \lambda^2 \Lambda^3 .
\eeq
Unless $\lambda \not= \ord{1}$, {\it we find stabilization
at $\Lambda \approx m$ using only a single condensate
and without reference to nonperturbative corrections
to the K\"ahler potential.}
Regardless of the value of $\lambda$,
the ``minimum'' is outside of the regime of validity
of the approximation \myref{thap} made above.  This is clear
from the far right-hand side of \myref{jhta}:
the ``correction'' must cancel against the leading
order to have $W'(S) \approx 0$.
Thus it is not possible
to draw any firm conclusions without pursuing the higher
order corrections in \myref{pwsr}.
Nevertheless it is interesting
that a nontrivial minimum does exist
for the truncated correction \myref{thap}.
In particular, we find it significant that
$|W'(S)|^2$ depends on {\it both}
$S+\Sb$ and $S-\Sb$.  Thus the axion can be stabilized
with just a single condensate.  {\it This}
qualitative result should continue to hold even
when the additional corrections in \myref{pwsr}
are taken into account.\footnote{At the final stages
of preparing this manuscript, we became aware of \cite{Escoda:2003fa},
where nontrivial vacua due to the DV corrections
are also studied.}

\section{Outlook}
\label{outl}
In rigid supersymmetry, we have arranged for
the equations of motion of the first-order
system to enforce {\it all} constraint equations
of the chiral and linear systems.
Thus we are assured that the two formulations
are in every way equivalent:  they have
equivalent equations of motion {\it and} constraint
equations.  This same approach may be applied to
locally supersymmetric extensions, which have
phenomenological applications to string-inspired
effective supergravity theories.  The sort of approach
taken here will connect the two formulations in
a way that is a faithful translation.

We have elucidated how the component fields of the
modified linear multiplet can accomodate the features
of a chiral multiplet with an axion potential.
In essence, the linear multiplet becomes a more
general sort of real multiplet when it couples
to the SYM instantons.  Its component field content
has a more general structure and modified interpretation.

We have suggested how the DV corrections can lead
to stabilization with a single condensate.  This should
come as no surprise.  We are not {\it forced} to integrate
out the adjoint chiral superfield $\Phi$ that appears
in \myref{uurq}.  If we leave the field in it has nontrivial
vacua and this additional condensate plays a role
in the stabilization of $S$.

It is interesting to consider the DV corrections in
the context of string-inspired effective supergravity.
Here it may be possible to achieve
a stable minimum where the DV corrections
are small ($\Lambda \ll m$), unlike the case above.
(For this to be true, however, it would seem that the
supergravity effects would have to dominate over
those that arise from the exchange of the adjoint
chiral superfield $\Phi$.)  In this
case the truncations made above become reasonable approximations
and the corrections may have some modest effects on,
for example, soft supersymmetry breaking operators in
the low energy effective theory.

We are currently investigating these and related issues.

\vspace{15pt}

\noindent
{\bf \large Acknowledgements} \vspace{5pt} \\
J.G.~would like to thank Erich Poppitz for numerous
helpful conversations, and the Michigan Center
for Theoretical Physics and Gordy Kane for hospitality during
initial stages of this research.  J.G.~received 
support from the National Science and Engineering 
Research Council of Canada and from the Ontario 
Premier's Research Excellence Award.

\end{document}